\documentclass[12pt]{article}
\usepackage{amssymb,amsfonts} 
\usepackage[dvips]{hyperref}
\def\rank{{\rm rank}}
\def\odd{{\rm odd}}
\def\Odd{{\rm Odd}}
\def\even{{\rm even}}
\def\Even{{\rm Even}}
\newtheorem{theorem}{Theorem} 
\newtheorem{corollary}{Corollary} 
\newtheorem{lemma}{Lemma} 
\newtheorem{statement}{Statement} 
\newtheorem{remark}{Remark} 

\newcommand\eq[1]{\begin{equation} #1 \end{equation}}
\newcommand\bm{\textbf}

\def\remark{\vspace{0.5em}\par\noindent\bf Remark~\refstepcounter{remark}\theremark.~\rm}

\def\Lemma{\vspace{0.5em}\par\noindent\bf Lemma~\refstepcounter{lemma}\thelemma~\it}
\def\statement{\vspace{0.5em}\par\noindent\bf Proposition~\refstepcounter{statement}\thestatement.~\it}
\def\corollary{\vspace{0.5em}\par\noindent\bf Corollary~\refstepcounter{corollary}\thecorollary.~\it}
\def\theorem{\vspace{0.5em}\par\noindent\bf Theorem~\refstepcounter{theorem}\thetheorem.~\it}
\def\qef{\vspace{0.5em}\rm\par}
\def\proof{\par{\bf Proof.~}}

\title{$Z_4$-Linear Perfect Codes%
\renewcommand{\thefootnote}{*}%
\footnote{Original Russian text was published in Diskretn. Anal. Issled. Oper., Ser.~1, 7(4):78-90, 2000.}
}
\author{\href{http://arxiv.org/find/grp_math/1/au:+Krotov_Denis/0/1/0/all/0/1}{D.\,S.\,Krotov}}
\begin{document}
\maketitle
\begin{abstract} For every  $n=2^k\geq 16$ there exist
exactly $\lfloor (k+1)/2\rfloor$ mutually nonequivalent
$Z_4$-linear extended perfect codes with distance $4$.
All these codes have different ranks.
\end{abstract}

\newcommand\bibi{\ddag}
Certain of known nonlinear binary codes such as Kerdock, Preparata, Goethals,
Delsarte-Goethals codes can be represented,
using some mapping $Z_4\to Z_2^2$
(in this paper, following \cite{HammonsOth:Z4_linearity}, we use the mapping
$0\to00$,
$1\to01$,
$2\to11$,
$3\to10$)
as linear codes over the alphabet $\{0,1,2,3\}$ with modulo $4$ operations
(see \cite{Nechaev:Kerdock_cyclic_form,KKMN:95,KuNe:92Gal,KuNe:94char2,HammonsOth:Z4_linearity}).
Codes represented in such a manner are called $Z_4$-linear.
In \cite{HammonsOth:Z4_linearity} it is shown that the extended Golay code and the
extended Hamming $(n,2^{n-\log_2n-1},4)$-codes
(of length $n$ and cardinality $2^{n-\log_2n-1}$, with distance $4$)
for every $n>16$ are not $Z_4$-linear. Also, in \cite{HammonsOth:Z4_linearity}
for every $n=2^k$ a $Z_4$-linear $(2,2^{n-\log_2n-1},4)$-code is described
(the codes $C^{0,r_2}$, in the notations of \S\,2, are presented as cyclic codes in \cite{HammonsOth:Z4_linearity}).
The goal of this work is a complete description of $Z_4$-linear perfect and extended perfect codes.

It is known \cite{ZL:1973,Tiet:1973} that there are no nontrivial perfect binary codes
except the Golay $(23,2^{12},7)$-code and the $(2^k-1,2^{2^k-k-1},3)$-codes.
The perfect $(23,2^{12},7)$-code is unique up to equivalence.
The linear (Hamming) $(2^k-1,2^{2^k-k-1},3)$-code is also unique for every $k$,
but for $n=2^k-1\geq 15$ there exist more than $2^{2^{(n+1)/2-k}}$
(for the last lower bound, see  \cite{Kro:quas_l_b})
nonlinear codes with the same parameters
(see, e.g., \cite{Sol:98constr,CHLL} for a survey of some constructions).
The class of all $(2^k-1,2^{2^k-k-1},3)$-codes is not described yet.

In this paper we show that not great, but increasing as $k\to\infty$,
number of extended perfect $(2^k,2^{2^k-k-1},4)$-codes can be represented
as linear codes over the ring $Z_4$.
In \S\,\ref{s:2}, in terms of check matrices, we define $\lfloor(\log_2n+1)/2\rfloor$
$Z_4$-linear extended perfect $(n,2^n/2n,4)$-codes.
In \S\,\ref{s:3} we show that the codes constructed are pairwise nonequivalent.
In \S\,\ref{s:4} we prove the nonexistence of $Z_4$-linear $(n,2^n/2n,4)$-codes
that are nonequivalent to the codes constructed.
In \S\,\ref{s:5} we propose an inductive way to construct the class of
$Z_4$-linear extended perfect codes.

So, all the $Z_4$-linear $(n,2^n/2n,4)$-codes
are described up to equivalence.
By the definition, codes of odd length cannot be $Z_4$-linear.
The length-$24$ Golay code, as noted above, is also non-$Z_4$-linear \cite{HammonsOth:Z4_linearity}.
Obviously, all trivial perfect and extended perfect binary codes
(the code from one all-zero word, the repetition $(n,2,n)$-code,
the all-parity-check $(n,2^{n-1},2)$-code,
the complete $(n,2^n,1)$-code) are $Z_4$-linear provided the length is even.
So, the problem of describing all $Z_4$-linear
perfect and extended perfect codes has got an exhaustive decision.

\subsection{Translator's remarks}\label{s:0}
In this section we briefly survey results closely related with the subject of this manuscript
but not cited in the original Russian-language paper.

As noted above, (non-extended) perfect distance $3$ code cannot be $Z_4$-linear.
Never\-the\-less, they can have a mixed additive $Z_2Z_4$ structure \cite{BorRif:1999}
(using an isometric mapping $Z_2^{k_1}Z_4^{k_2}\to Z_2^{k_1+2k_2}$, one can construct
binary codes from group codes in $Z_2^{k_1}Z_4^{k_2}$).
In \cite{BorRif:1999},
all such perfect codes of length $n=2^k-1$ are characterized;
it turns out that there are exactly
$\lfloor k/2 \rfloor+1$ such codes, up to equivalence.

The ranks and dimensions of kernels of the additive perfect and extended perfect binary codes
(including the class considered in this paper)
are calculated in \cite{PheRif:2002} and \cite{BorPheRif:2003}.

In \cite{Kro:WCC'2001:Z4_Had_Perf}, the codes whose $Z_4$-preimage
is dual to the preimage of some $Z_4$-linear
extended perfect code are considered (by MacWilliams-type theorems,
such a code has the parameters of the first order Reed-Muller code $RM(1,k)$, or a Hadamard code);
the number of such codes of length $2^k$ is $\lfloor (k-1)/2 \rfloor$
(in the notation of this paper,
the codes $\phi({\cal C}^{0,r_2*})$ and $\phi({\cal C}^{1,r_2-2*})$ are equivalent).
All the additive
codes with parameters of $RM(1,k)$, including the $Z_4$-linear case,
and their ranks and kernels are characterized in \cite{PheRifVil:2006}.
The series of $Z_4$-linear extended perfect and Hadamard codes can be generalized
to the series of codes with the parameters of Reed-Muller codes $RM(o,k)$ for all orders $o$,
$0\leq o\leq k$, see \cite{Sol:2007RM,PRS:2007Plot} (recall that extended perfect
and Hadamard codes of length $2^k$ have the parameters of $RM(k-2,k)$ and $RM(1,k)$, respectively).

The construction of co-$Z_{2^k}$-linear extended perfect codes
and $Z_{2^k}$-linear Hadamard codes (where the meaning of $k$ is not the same as above)
presented in \cite{Kro:2007Z2k} generalizes the construction
of this paper and \cite{Kro:WCC'2001:Z4_Had_Perf} to the $Z_8$, $Z_{16}$, $Z_{32}$,\ldots cases.

Another generalization of $Z_4$-linear and additive codes is transitive codes,
when the stabilizer of the code in the isometry group of the space acts transitively
on the codewords (for each two codewords $\bar x$,  $\bar y$ there is an element $\pi$ of
the stabilizer such that $\pi(\bar x)=\bar y$). Wide classes of transitive perfect binary codes
are constructed in \cite{Sol:2005transitive,Pot:trans}. As shown in \cite{Pot:trans},
the number of such codes grows at least exponentially with respect to the square root of the length.

The recurrent construction in \S\,\ref{s:induct} originates
from the Mollard construction \cite{Mollard}.

\section{Main concepts and notations}\label{s:1}

 Denote the set of all binary words of length $n$ by $E^n$ .
 The {\it Hamming distance} $d(x,y)$ between two words $x,y\in E^n$ is
 the number of positions in which $x$ and $y$ differ.
 A set $C\subset E^n$
 is called a binary {\it $(n,K,d)$-code}
 if $|C|=K$ and the Hamming distance between any two different
 words in $C$ is not less than $d$.
 A code $C$ is called {\it linear}
 if it is closed under the modulo $2$ addition.

 A code $C$ with parameters $(n,K,2\rho+1)$ is called  {\it perfect} if
 the distance from any word of $E^n$ to $C$ does not exceed $\rho$.
 An $(n,K,2\rho+2)$-code is called {\it extended perfect}
 if removing the last symbol from every code word
 results in a perfect $(n-1,K,2\rho+1)$-code.
 An $(n,K,4)$-code is extended perfect if and only if $K=2^n/2n$.

By $Z_4^n$ we denote the set of length-$n$ words over the alphabet $Z_4=\{0,1,2,3\}$
with the modulo $4$ addition and multiplication by a constant.
We will say that a word $c\in Z_4^n$ has the {\it mixture} $1^{n_1}2^{n_2}3^{n_3}$
if $c$ contains $n_1$ ones, $n_2$ twos, $n_3$ threes and $n-n_1-n_2-n_3$ zeros
placed in an arbitrary order.
An additive subgroup of $Z_4^n$ will be called a {\it quaternary code}.
Two quaternary codes are called {\it equivalent} if
one can be obtained from the other by a coordinate permutation
and/or changing the sign in some coordinates.
At that, if we use only a coordinate permutation, then
the codes are {\it permutably equivalent}.

The {\it Lee weight} $wt_L(a)$ of a word $a$ from $Z_4^n$
is the usual (over $Z$) sum of Lee weights of all coordinates of $a$,
where $wt_L(0)=0$, $wt_L(1)=wt_L(3)=1$, and $wt_L(2)=2$.
This weight function defines the {\it Lee metric}
$d_L(a,b)=wt_L(b-a)$
on $Z_4^n$.
A quaternary code
${\cal C}\subset Z_4^n$ is called a {\it  quaternary distance-$d$ code of length $n$}
or an {\it $(n,{|\cal C|},d)_4$-code} if $d_L(a,b)\geq d$ for any different
$a,b\in\cal C$, which is equivalent to $wt_L(a)\geq d$ for any nonzero $a\in{\cal C}$.

Any quaternary code $\cal C$ can be defined by a {\it generating matrix} of form
\eq{G=\left[\matrix{G_1\cr 2G_2}\right]\label{gen}}
where $G_1$ is a $Z_4$-matrix of size $k_1\times n$,
$G_2$  is a $Z_2$-matrix of size $k_2\times n$,
$|{\cal C}|=2^{2k_1+k_2}$, and every word $c$ from $\cal C$ can be represented as
$$c=(v_1,v_2)\left[\matrix{G_1\cr 2G_2}\right]\mbox{ (mod 4)},
\qquad v_1\in Z_4^{k_1},\quad v_2\in Z_2^{k_2}.$$

The code $\cal C$ defined by the generating matrix (\ref{gen})
is an  elementary Abelian group of type $4^{k_1}2^{k_2}$.
We will indicate this as follows: $|{\cal C}|=4^{k_1}2^{k_2}$.

Words $x=(x_0,\ldots,x_{n-1})$ and
$x'=(x'_0,\ldots,x'_{n-1})$ from $Z_4^n$ (from $E^n$)
are said to be {\it dual}, i.e. $x\perp x'$,
if $x_0x'_0+\ldots+x_{n-1}x'_{n-1}=0\mbox{ (mod\,$4$)}$
(respectively, (mod\,$2$)).
The duality relation is naturally extended to the duality
of a word and a set of words and to the duality of two sets of words from $Z_4^n$ (from $E^n$).

A quaternary code ${\cal C}$ of type $4^{k_1}2^{k_2}$ can be described by
a {\it check matrix}
$$A=\left[\matrix{A_1\cr 2A_2}\right]\label{pro}$$
by the relation
$$Ac^T=0 \mbox{ for any }c\in{\cal C},$$
where $A_1$ is a $Z_4$-matrix of size $(n-k_1-k_2)\times n$ and
$A_2$ is a $Z_2$-matrix of size $k_2\times n$. The matrix $A$ is generating
for the quaternary code ${\cal C}^*$ that is {\it dual} to ${\cal C}$;
${\cal C}^*$ can be alternatively defined as the set of words that are dual to ${\cal C}$.

Let us define two maps $\beta(c)$ and $\gamma(c)$ from $Z_4$ to $Z_2=\{0,1\}$:
$$\matrix{c & \beta(c) & \gamma(c) \cr
          0&0&0\cr
          1&0&1\cr
          2&1&1\cr
          3&1&0},$$
and let they be extended to maps from $Z_4^n$ to $Z_2^n$ by coordinates.
The {\it Gray map} $\phi: Z_4^n\to E^{2n}$ is defined
by $$\phi(c)=(\beta(c),\gamma(c)),\quad c\in Z_4^n$$
(so, $i$th coordinate of $c$ corresponds to $i$th and $(i+n)$th
binary coordinates of $\phi(c)$).
Applying $\phi(\cdot)$ to every code word,
to arbitrary quaternary code we can assign a binary code
of twice length and the same cardinality.
Following \cite{HammonsOth:Z4_linearity}, we will denote quaternary codes
by calligraphic letters, and the corresponding binary codes,
by usual latin letters, e.g.,
$C=\phi({\cal C})$,
$B=\phi({\cal B})$,
$C^{2,3}=\phi({\cal C}^{2,3})$.
The binary code $C$ obtained by applying the Gray map
to all the words of some quaternary code $\cal C$, and
all the codes that can be obtained from $C$ by a coordinate permutation
are called {\it $Z_4$-linear}.

Two binary codes $C$ and $C'$ of length $n$ are called {\it equivalent}
if there exist a word $y$ from $E^n$ and a coordinate permutation $\pi$
such that $C=\pi(C'\oplus y)$.
If quaternary codes $\cal C$ and ${\cal C}'$ are equivalent,
then the corresponding binary codes $C$ and $C'$ are also equivalent
(changing the sign in the $i$th coordinate of length-$n$ code $\cal C$ corresponds
to the transposition $(i,i+n)$ of the coordinates of $C$).

It follows directly from the definitions
of the Hamming $d(\cdot,\cdot)$ and Lee $d_L(\cdot,\cdot)$ metrics
and the Gray map $\phi(\cdot)$ that
$$d_L(a,b)=d(\phi(a),\phi(b)),\quad a,b\in Z_4^n.$$
So, we have the following:

\Lemma {\rm\cite{HammonsOth:Z4_linearity}.} \label{thgr}
The map $\phi$ is an isometry between
the spaces $Z_4^n$ with the Lee metric
and $E^{2n}$ with the Hamming metric.
\qef

An $(n,4^n/4n,4)_4$-code will be called a {\it perfect} quaternary code.
As follows from Lemma~\ref{thgr}, a quaternary code $\cal C$ is perfect
if and only if $C$ is an extended perfect binary code with distance $4$.


\section{A construction of $Z_4$-linear extended perfect codes}\label{s:2}

Let $r_1$ and $r_2$ be nonnegative integers.
Let us compose the matrix $A^{r_1,r_2}$
from all different columns of type $z^T$,
$z\in \{1\}\times\{0,1,2,3\}^{r_1}\times\{0,2\}^{r_2}$ ordered lexicographically.
For example,
$$A^{0,0}=\left[\matrix{1}\right],\quad
    A^{0,1}=\left[\matrix{11\cr02}\right],$$
$$A^{1,0}=\left[\matrix{1111\cr0123}\right], \quad
    A^{0,2}=\left[\matrix{1111\cr0022\cr0202}\right],$$
$$A^{1,1}=\left[\matrix{11\,11\,11\,11\cr00\,11\,22\,33\cr02\,02\,02\,02}\right], \quad
    A^{0,3}=\left[\matrix{11\,11\,11\,11\cr
                          00\,00\,22\,22\cr
                          00\,22\,00\,22\cr
                          02\,02\,02\,02}\right],$$
$$A^{2,0}=\left[\matrix{1111\,1111\,1111\,1111\cr
                           0000\,1111\,2222\,3333\cr
                           0123\,0123\,0123\,0123}\right].$$

\theorem\label{th4_1} The quaternary code
$${\cal C}^{r_1,r_2}=\{c\in Z_4^{2^{2r_1+r_2}}:A^{r_1,r_2}c^T=0\}$$
is perfect.
\qef

\proof
The length $n$ of ${\cal C}^{r_1,r_2}$ equals $4^{r_1}2^{r_2}$, i.e., the number of elements
in $Z_4^{r_1}\times Z_2^{r_2}$.
The code ${\cal C}^{r_1,r_2*}$ with the generating matrix $A^{r_1,r_2}$
has type $4^{r_1+1}2^{r_2}$.
Therefore, $|C^{r_1,r_2}|=4^n/|C^{r_1,r_2*}|=4^{n-r_1-r_2-1}2^{r_2}=4^n/4n$.

Let us show that the weight of any nonzero word from ${\cal C}^{r_1,r_2}$ is not less than $4$.
The words with the mixtures $1$, $3$, $2$, $1^2$, $3^2$, $12$, $23$,
$1^3$, $1^23$, $13^2$, $3^3$
contradict to the first row of the matrix $A^{r_1,r_2}$.
A word of mixture $13$ cannot belong to ${\cal C}^{r_1,r_2}$ because
this would mean that the difference between some two columns of
$A^{r_1,r_2}$ equals zero, which means coincidence of these columns
and contradicts to the definition of $A^{r_1,r_2}$.
Theorem~\ref{th4_1} is proved.


\section{Pairwise nonequivalence of the constructed codes}\label{s:3}

Two $Z_4$-matrices $A$ and $A'$ with the same number of columns will be called
 \em equivalent \em if
every row of $A$ is a linear combination of rows of $A'$
and vice versa,
every row of $A'$ is a linear combination of rows of $A$.

Below, we define the functions
 $\Even$, $\Odd$, $\even$, and $\odd$,
which will be used to prove statements by induction.

Let $n$ be even.
Assume that $a_0, a_1,...,a_{n-1}$ are the columns of a matrix
$A=(a_0,a_1,\ldots,a_{n-1})$; then by $\Even(A)$ and $\Odd(A)$ we will
denote the matrices $(a_0,a_2,\linebreak[1]\ldots,a_{n-2})$ and $(a_1,a_3,\ldots,a_{n-1})$,
which are composed from the even and the odd columns of $A$
(i.e., the columns $a_i$ with odd/even indexes $i$), respectively.
Similarly define $\Even(x)$ and $\Odd(x)$ for a word
$x=(x_0,\ldots,x_{n-1})$ from $Z_4^n$ or
$E^n$.

\statement \label{AA}\\
{\rm a)} For any $r_1\geq 0$ and $r_2>0$
the matrices $\Even(A^{r_1,r_2})$ and $\Odd(A^{r_1,r_2})$ are equivalent
to the matrix $A^{r_1,r_2-1}$.\\
{\rm b)} For any  $r_1>0$
the matrices $\Even(A^{r_1,0})$ and $\Odd(A^{r_1,0})$ are equivalent to $A^{r_1-1,1}$.
\qef

\proof
a) By the definition, the matrix $A^{r_1,r_2-1}$ is obtained from
$\Even(A^{r_1,r_2})$ or $\Odd(A^{r_1,r_2})$ by removing the last row.
The last row of $\Even(A^{r_1,r_2})$ consists of zeros;
the last row of $\Odd(A^{r_1,r_2})$ consists of twos and, thus,
is equal to the first row of $A^{r_1,r_2-1}$ multiplied by $2$.

b) The matrix $A^{r_1-1,1}$ coincides with $\Even(A^{r_1,0})$
and can be obtained from $\Odd(A^{r_1,0})$ by subtracting the first row from the last,
which consists of $1$s and $3$s.
Proposition~\ref{AA} is proved.
\def\qed{\vspace{0.5em}}
\qed

For ${\cal C}\subset Z_4^n$ we denote
$$\even({\cal C})\stackrel{\rm\scriptscriptstyle def}=
\{(c_0,c_2,\ldots,c_{n-2})\in Z_4^{n/2}\,|\,
(c_0,0,c_2,0,\ldots,c_{n-2},0)\in {\cal C}\},$$
$$\odd({\cal C})\stackrel{\rm\scriptscriptstyle def}=
\{(c_1,c_3,\ldots,c_{n-1})\in Z_4^{n/2}\,|\,
(0,c_1,0,c_3,\ldots,0,c_{n-1})\in {\cal C}\}.$$
Similarly we define $\even(C)$ and $\odd(C)$ for $C\subset E^n$.

The following three propositions are straightforward from the definitions.

\statement\label{42} Let ${\cal C}\subset Z_4^n$ and ${\cal B}\subset Z_4^{n/2}$ be quaternary codes.
  Then\\
{\rm a)} ${\cal B}=\even({\cal C})$ if and only if $B=\even(C),$\\
{\rm b)} ${\cal B}=\odd({\cal C})$ if and only if $B=\odd(C)$.

\statement\label{x4}
Let $C\subset E^n$, $y\in E^n$, and $y\perp C$.
Then $\Even(y)\perp \even(C)$ and
 $\Odd(y)\perp \odd(C)$.

\statement\label{A4}
Let $A$ be a check matrix of a quaternary code $\cal C$.
Then $\Even(A)$ is a check matrix of the code $\even({\cal C});$
 $\Odd(A)$ is a check matrix of the code $\odd({\cal C})$.
\qef

From Propositions~\ref{AA}, \ref{A4}, and \ref{42}, we conclude the following:
\corollary\label{CCC}\\
{\rm a)}
$\even(C^{r_1,r_2})=\odd(C^{r_1,r_2})=C^{r_1,r_2-1}$ for any $r_1\geq 0$ and $r_2>0$.\\
{\rm b)}
$\even(C^{r_1,0})=\odd(C^{r_1,0})=C^{r_1-1,1}$ for any $r_1>0$.
\qef

The maximum number of linearly independent vectors in a binary code $C$ as called the rank
of $C$ and denoted by $\rank(C)$.
The rank of a code $C$ equals to the length of $C$ minus the maximum number of linearly independent vectors
that are dual to $C$.
If two codes containing the all-zero word have different ranks, then they are nonequivalent.

We call a binary word
$y=(y_0,\ldots,y_{n-1})$ of even length $n$ {\it repetitive} if for all $i\in \{0,\ldots,n/2-1\}$
it holds $y_i=y_{i+n/2}$.
In other words, $y$ is repetitive if and only if $\phi^{-1}(y)\in\{0,2\}^{n/2}$.
Obviously, the sum of   repetitive words is repetitive.

\statement\label{op4} If $x,x'\in\{0,2\}^n\subset Z_4^n$,
then $\phi(x+x')=\phi(x)\oplus\phi(x')$.
\qef

\proof
Since the addition of words from $Z_4^n$ and the addition of words from $E^{2n}$ are defined
coordinatewise,
it is enough only to check that
$\phi(x_0+x'_0)=\phi(x_0)\oplus\phi(x'_0)$ for $x_0,x'_0\in\{0,2\}$, which is straightforward.
Proposition~\ref{op4} is proved.

\statement\label{perp02} Let $\cal C$ be a quaternary code of length $n$;
and let $x\in\{0,2\}^n$.
Then $x \perp {\cal C}$ and $\phi(x)\perp C$ are equivalent.
\qef

\proof
We have to show that
 $x \perp c$ is equivalent to $\phi(x)\perp\phi(c)$ for an arbitrary $c\in\cal C$.
Let $k$ be the number of $2$s in $x$; and let $i_1,\ldots,i_k$ be the numbers of positions
in which $x$ contains $2$. Then $x\perp c$ means that
$\sum_{j=1}^k2c_{i_j}=0\mbox{ (mod\,$4$)}$ and is equivalent
to the evenness of the sum of all $c_{i_j}$, $j=1,\ldots,k$, which is equivalent
to the evenness of the sum of all $\beta(c_{i_j})$ and $\gamma(c_{i_j})$, $j=1,\ldots,k$,
which, in its turn, is equivalent to the relations
$\sum_{j=1}^k (\phi(c)_{i_j}\oplus\phi(c)_{i_j+n})=0\mbox{ (mod\,$2$)}$, i.e.,
$\phi(x)\perp\phi(c)$.
Proposition~\ref{perp02} is proved.

\statement\label{ort2}
For any integer $r_1\geq 0$, $r_2\geq 0$ the dimension of the subspace
of repetitive words from $E^{2^{2r_1+r_2+1}}$ that are dual to $C^{r_1,r_2}$
equals $r_1+r_2+1$.
\qef

\proof
Let $a_0,a_1,\ldots,a_{r_1+r_2}$ be, respectively, the
first, the second, \ldots, the $(r_1+r_2+1)$th rows of the matrix $A^{r_1,r_2}$.
Then the words
$2a_0,\linebreak[1] 2a_1,\linebreak[1] \ldots,\linebreak[1] 2a_{r_1},\linebreak[1] a_{r_1+1},\linebreak[1] \ldots,\linebreak[1] a_{r_1+r_2}$
consist of $0$s and $2$s; so,
by Proposition~\ref{perp02},
the repetitive linearly independent words
\begin{equation}
{\label{words4}\phi(2a_0),\phi(2a_1),\ldots,\phi(2a_{r_1}),\phi(a_{r_1+1}),
\ldots,\phi(a_{r_1+r_2})}
\end{equation}
are dual to $C^{r_1,r_2}$.

On the other hand, if $y$ is a repetitive word that is dual to $C^{r_1,r_2}$,
then the word $\phi^{-1}(y)\in\{0,2\}^{2^{2r_1+r_2}}$
is dual to ${\cal C}^{r_1,r_2}$. Consequently, $\phi^{-1}(y)$ is a linear combination
of rows of $A^{r_1,r_2}$. Since
$2\phi^{-1}(y)$ is the all-zero word, we see that the coefficients
at the $r_1+1$ rows in this linear combination are even.
So, $\phi^{-1}(y)$ is a linear combination of the words
$2a_0,2a_1,\ldots,2a_{r_1},a_{r_1+1},\ldots,a_{r_1+r_2}$,
and, by Proposition~\ref{op4}, the word $y$ is a linear combination of the words (\ref{words4}).
Proposition~\ref{ort2} is proved.

\corollary\label{r1r2}
For any integer $r_1\geq 0$, $r_2\geq 0$ it holds
$$\rank(C^{r_1,r_2})\leq n-r_1-r_2-1,$$
where $n=2^{2r_1+r_2+1}$ is the length of $C^{r_1,r_2}$.

\statement\label{rank0} For any integer $r_2\geq 4$ it holds
$$\rank(C^{0,r_2})=2^{r_2+1}-r_2-1=n-\log_2n,$$
where $n=2^{r_2+1}$ is the length of $C^{0,r_2}$.
\qef

\proof
It is shown in \cite{HammonsOth:Z4_linearity} that the linear extended perfect Hamming codes of length more than $16$
are not $Z_4$-linear. Consequently,
for $r_2\geq 4$ the code $C^{0,r_2}$ is nonlinear, and its rank is greater than
$n-\log_2n-1$ (the dimension of the Hamming code).
But, by Corollary~\ref{r1r2}, the rank of $C^{0,r_2}$
does not exceed $n-\log_2n$.
Proposition~\ref{rank0} is proved.

\remark Proposition~\ref{rank0} can be proved in the same manner as
Corollary~\ref{rank} below, by induction, after establishing the nonlinearity of $C^{0,4}$.
This way do not use the
``non-$Z_4$-linearity'' of the Hamming codes, and
this ``non-$Z_4$-linearity'' can be independently derived as a corollary of
the nonlinearity of the $Z_4$-linear codes $C^{r_1,r_2}$.

\statement\label{C16} The rank of $C^{1,1}$ is $13$.
\qef

\proof By Corollary~\ref{r1r2}, $\rank(C^{1,1})\leq 13$.
Let us list $13$ linearly independent vectors from $C^{1,1}$:
\begin{eqnarray*}
b_1 =\phi(2200\,0000)=1100\,0000\ 1100\,0000\nonumber\\
b_2 =\phi(0000\,2200)=0000\,1100\ 0000\,1100\nonumber\\
b_3 =\phi(2000\,2000)=1000\,1000\ 1000\,1000\nonumber\\
b_4 =\phi(1100\,1100)=0000\,0000\ 1100\,1100\nonumber\\
b_5 =\phi(0022\,0000)=0011\,0000\ 0011\,0000\nonumber\\
b_6 =\phi(0000\,0022)=0000\,0011\ 0000\,0011\nonumber\\
b_7 =\phi(0020\,0020)=0010\,0010\ 0010\,0010\nonumber\\
b_8 =\phi(0011\,0011)=0000\,0000\ 0011\,0011\nonumber\\
b_9 =\phi(0000\,1313)=0000\,0101\ 0000\,1010\nonumber\\
b_{10}=\phi(0101\,0303)=0000\,0101\ 0101\,0000\nonumber\\
b_{11}=\phi(0101\,3030)=0000\,1010\ 0101\,0000\nonumber\\
b_{12}=\phi(1000\,0111)=0000\,0000\ 1000\,0111\nonumber\\
b_{13}=\phi(0100\,0102)=0000\,0001\ 0100\,0101
\end{eqnarray*}
The collection $b_1,\ldots,b_{11}$ is a basis of the Hamming code with the check matrix
\[ B=\left[ \matrix{ 1111\,1111\,1111\,1111\cr
                     0000\,0000\,1111\,1111\cr
                     0000\,1111\,0000\,1111\cr
                     0011\,0011\,0011\,0011\cr
                     0101\,0101\,0101\,0101   } \right]. \]
The vector $b_{12}$ is dual to all rows of $B$ except the third one;
consequently, $b_{12}$ is linearly independent of $b_1,\ldots,b_{11}$.
The vector  $b_{13}$ is not dual to the second row of $B$,
consequently, $b_{13}$ is linearly independent of $b_1,\ldots,b_{12}$.
Proposition~\ref{C16} is proved.

\corollary \label{C11} All the words of $E^{16}$ that are dual to $C^{1,1}$ are repetitive.
\qef

\proof
Otherwise, there exist at least four linearly independent words that are dual to
 $C^{1,1}$:
three  repetitive (Proposition~\ref{ort2}) and one non repetitive.
This means that $\rank(C^{1,1})\leq 16-4=12$, which contradicts to
Proposition~\ref{C16}.
Corollary~\ref{C11} is proved.

\statement\label{ind}
Let $r_1\geq 1$ and $r_2\geq 0$ be integers satisfying $2r_1+r_2\geq 3$.
Then all the words of $E^{2^{2r_1+r_2+1}}$ that are dual to $C^{r_1,r_2}$ are repetitive.
\qef

\proof We will argue by induction on $r=2r_1+r_2$.

By Corollary~\ref{C11}, the statement holds for $r=3$.

Assume that it holds for $r=k-1\geq 3$. Let $2r_1+r_2=k$
and $y\perp C^{r_1,r_2}$. Then, by Proposition~\ref{x4}, we have
$\Even(y)\perp \even(C^{r_1,r_2})$
and $\Odd(y)\perp \odd(C^{r_1,r_2})$. By Corollary~\ref{CCC},
 $\odd(C^{r_1,r_2})=\even(C^{r_1,r_2})=C^{r_1,r_2-1}$ for $r_2>0$ and
 $\odd(C^{r_1,r_2})=\even(C^{r_1,r_2})=C^{r_1-1,r_2+1}$ for $r_2=0$;
 thus, by the inductive assumption,
the words $\Even(y)$ and $\Odd(y)$ are repetitive,
and $y$ is also repetitive by the Definition.
Pro\-po\-si\-tion~\ref{ind} is proved.
\qed

Propositions~\ref{ind} and~\ref{ort2} yield the following:
\corollary\label{rank} Let $r_1\geq 1$, $r_2\geq 0$ be integers satisfying
$2r_1+r_2\geq 3$. Then $$\rank(C^{r_1,r_2})=2^{2r_1+r_2}-r_1-r_2-1.$$

\theorem\label{th4_2} Let $2r_1+r_2=2r'_1+r'_2\geq 3$;
then the codes $C^{r_1,r_2}$ and $C^{r'_1,r'_2}$ are equivalent if and only if $r_1=r'_1$.
\qef

\proof
In the case  $r=2r_1+r_2=2r'_1+r'_2\geq 4$, by Proposition~\ref{rank0}
and Corollary~\ref{rank}, the codes
$C^{r_1,r_2}$ and $C^{r'_1,r'_2}$ have the ranks
$2^{r}-r+r_1-1$ and
$2^{r}-r+r'_1-1$, respectively.
If $r_1\neq r'_1$, then the ranks are different, and the codes are nonequivalent.

In the case $r=3$ we have to show that $C^{0,3}$ and $C^{1,1}$ are nonequivalent.
This is true because $\rank(C^{0,3})\leq 12$ and $\rank(C^{1,1})=13$
(See Corollary~\ref{r1r2} and Proposition~\ref{C16}).
Theorem~\ref{th4_2} is proved.
\remark In fact, the code $C^{0,3}$ is linear and its rank equals $11$.


\section{The nonexistence of $(n,4^n/4n,4)_4$-codes
that are nonequivalent to the constructed codes} \label{s:4}

In the further investigation, the following two auxiliary statements are useful.
\statement\label{E_V}
If $C$ is an extended perfect distance-$4$ code of length $n$
and $x$ is a binary word dual to $C$, then $wt(x)=0$, $wt(x)=n/2$, or $wt(x)=n$.
\qef

This statement is equivalent to the fact that
a perfect binary distance-$3$ code is dual only to weight-$(n+1)/2$ and weight-$0$ vectors
(see, e.g., \cite{EV:94}).

\statement \label{0n}
If $n$ is a power of two and $D$ is a linear binary code of length $n$
whose all nonzero words have weight $n/2$, then all the words of $D$ have
a common zero coordinate.
\qef

\proof
We will prove the statement by induction..
In the case $n=2$ the statement is obvious
(we can also consider the trivial case $n=1$ as the induction base).

Assume that the statement holds for $n=m/2$.
Let us show that it is true for $n=m$.
Without loss of generality assume that $D$ contains the word $b=(0,1,0,1,\ldots,0,1)$
with zeros in even coordinates and ones, in odd.
Any other nonzero word $b'$ from $D$ contains
$m/4$ ones in even coordinates, and the same number, in odd,
because $wt(b\oplus b')=m/2$. Consequently, all the nonzero words of
the code $D'=\{\Even(d)|d\in D\}$ of length $m/2$ have the weight $m/4$.
By the inductive assumption, all the words of $D'$ contain zero
in some common $i$th coordinate, $0\leq i\leq m/2-1$;
it follows that
all the words of  $D$ contain zero in the $2i$th coordinate.
Proposition~\ref{0n} is proved.

\theorem\label{th4_3}
Let ${\cal C}\subset Z_4^n$ be a $(n,4^n/4n,4)_4$-code,
and let $|{\cal C}|=4^{n-r_0-r_2}2^{r_2}$.
Then $r_0>0$ and $\cal C$ is equivalent to ${\cal C}^{r_0-1,r_2}$.
\qef

\proof
Since $4^n/4n=4^{n-r_0-r_2}2^{r_2}$, we have
\begin{equation}{n=2^{2r_0+r_2-2}.\label{mosh}}\end{equation}

Let a matrix $A$ of size $(r_0+r_2)\times n$
be a check matrix of $\cal C$, and let
$a^0, a^1,\ldots,a^{r_0+r_2-1}$ be its rows, where
$a^{r_0}\ldots,a^{r_0+r_2-1}\in \{0,2\}^n$.
Consider the repetitive words
$b^i=\phi(2a^i)$, $i=0,\ldots,r_0-1$, which are dual do $C$ by
Proposition~\ref{perp02}.
Let $D$ be the linear span of the set of words $\{b^i\}_{i=0}^{r_0-1}$.
By Proposition~\ref{E_V}, the linear code $D$ of length $2n$
consists of words of weights $0$, $n$, and $2n$.
Let $\overline 1\in E^{2n}$ be the all-one word (of weight $2n$).
Let us show that $\overline 1\in D$.

Suppose, by contradiction, that $D$ contains only words of weights $n$ and $0$.
Then, by Proposition~\ref{0n}, there exists $j$, $0\leq j\leq 2n-1$,
such that $d_j=0$ for every $d=(d_0,\ldots,d_{2n-1})\in D$.
Since all the words of $D$ are repetitive,
we also have $d_{j+n\,({\rm mod}\,2n)}=0$ for every $d\in D$;
consequently,
$\phi^{-1}(d)_{j'}=0$, where $j'=j\mbox{ (mod $n$)})$.
In particular, $2a^i_{j'}=0$ for every $i=0,\ldots,r_0-1$.
This means that the $j$th column of the matrix $A$ consists of zeros and twos,
which implies that $\cal C$ contains a weight-$2$ word
(with $2$ in $j'$th coordinate and zeros in the others).
We get a contradiction with the code distance $4$ of $\cal C$.

So, $\overline 1\in D$, and there are coefficients
$\alpha_0,\ldots,\alpha_{r_0-1}\in \{0,1\}$ such that
\begin{equation}{\label{e4_1}
\alpha_0b^0\oplus\ldots\oplus\alpha_{r_0-1}b^{r_0-1}=\overline 1.
}\end{equation}
This implies that $r_0\geq 1$.
Without loss of generality we assume $\alpha_0=1$
(otherwise we can permute the rows of $A$ in such a way that
the coefficient at $b_0$ in (\ref{e4_1}) will be nonzero).
Consider the matrix $A'$ obtained from $A$ by replacing the first row
$a^0$ by
$$a'^0=\alpha_0a^0+\ldots+\alpha_{r_0-1}a^{r_0-1} \mbox{ (mod $4$)}.$$
Since $\alpha_0=1$, we can represent $a^0$ as a linear combination of
$a'^0,a^1,\ldots,a^{r_0-1}$; so, the matrices $A$ and $A'$ are equivalent.

It follows from (\ref{e4_1}) that
$$2a'^0=\alpha_02a^0+\ldots+\alpha_{r_0-1}2a^{r_0-1}=2\cdot\overline1\mbox{ (mod $4$)},$$
i.e., $a'^0$, the first row of $A'$, consists of $1$s and $3$s.
Let $A''$ be obtained from $A'$ by changing the sign in the columns that have $3$
in the first position.
$A''$ is a check matrix of a quaternary code $\cal C''$, which is equivalent to $\cal C$
(can be obtained from $C$ by changing the sign in the corresponding coordinates).
Furthermore, the first row of $A''$ consists of $1$s; and the last $r_2$ rows consist of
$0$s and $2$s.
If $A''$ has two equal columns, say $j$th and $j'$th,
then $C''$ contains the weight-$2$ word with $1$ in the $j$th coordinate,
$3$ ($3=-1\mbox{ (mod $4$)}$) in  $j'$th, and zeros in the other coordinates.
This contradicts to the code distance $4$.
So, all columns of $A''$ are distinct; as follows from (\ref{mosh}),
$A''$ consists of all different columns of height $r_0+r_2$
with $1$ in the first position, $0$s and/or $2$s in the last $r_2$ positions,
and arbitrary numbers from $\{0,1,2,3\}$ in the other $r_0-1$ positions.
Ordering the columns lexicographically,  we obtain $A^{r_0-1,r_2}$;
applying the corresponding coordinate permutation to the words of $\cal C''$,
we obtain ${\cal C}^{r_0-1,r_2}$. So,
the code $\cal C''$ and, thus, the code $\cal C$ are equivalent to ${\cal C}^{r_0-1,r_2}$.
Theorem~\ref{th4_3} is proved.

\theorem\label{z4main}
Let $n=2^k\geq 16$; then
there exist exactly $\lfloor(\log_2n+1)/2\rfloor$
pairwise nonequivalent $Z_4$-linear extended perfect distance-$4$ codes of length $n$.
\qef

\proof
There are $\lfloor(\log_2n+1)/2\rfloor$ ways to represent $n$
as $n=2^{2r_1+r_2+1}$ with integers $r_1\geq 0$ and $r_2\geq 0$.
By Theorem~\ref{th4_2},
$\bm C=\{C^{r_1,\log_2n-2r_1-1}\}_{r_1=0}^{\left\lfloor(\log_2n-1)/2\right\rfloor}$
is a set of pairwise nonequivalent codes.
By Theorem~\ref{th4_3}, any $Z_4$-linear $(n,2^n/2n,4)$-code
is equivalent to one of the codes from $\bm C$.
Theorem~\ref{z4main} is proves.


\section{An inductive construction of the codes ${\cal C}^{r_1,r_2}$}\label{s:induct}\label{s:5}

Let $n'=4^{r'_1}2^{r'_2}$ and $n''=4^{r''_1}2^{r''_2}$ be powers of $2$ and

$$c=(c_{0,0},c_{0,1},...,c_{0,n''-1},c_{1,0},c_{1,1},\ldots,c_{n'-1,n''-1})\in Z_4^{n'n''}.$$
Denote
$$ p'(c)=\left(\sum_{j=0}^{n''}c_{0,j},
\sum_{j=0}^{n''}c_{1,j},\ldots,
\sum_{j=0}^{n''}c_{n'-1,j}\right)\quad \mbox{(mod $4$)},$$
$$ p''(c)=\left(\sum_{i=0}^{n'}c_{i,0},
\sum_{i=0}^{n'}c_{i,1},\ldots,
\sum_{i=0}^{n'}c_{i,n''-1}\right)\quad \mbox{(mod $4$)}.$$
(If we represent $c$ as a matrix of size $n'\times n''$, then
$p'$ is a sum of columns and $p''$ is a sum or rows of this matrix.)

Let $\cal C'$ be a quaternary code with a check matrix $A'$
that is permutably equivalent to ${\cal C}^{r'_1,r'_2}$,
and let
$\cal C''$ be a quaternary code with a check matrix $A''$,
that is permutably equivalent to ${\cal C}^{r''_1,r''_2}$.
Let $n=n'n''$, $r_1=r'_1+r''_1$ and $r_2=r'_2+r''_2$.
\theorem The set
 \begin{equation}\label{indC}
 {\cal C}=\{c\in Z_4^{n'n''}\,|\, p'(c)\in {\cal C'},\ p''(c)\in {\cal C''}\}
 \end{equation}
 is a quaternary $(n,4^n/4n,4)_4$-code which is permutably
 equivalent to ${\cal C}^{r_1,r_2}$.
\qef

The linearity of $\cal C$ over $Z_4$ is obvious;
the code distance and the cardinality are calculated in \cite{Kro:00univers}
for a more general construction;
the type of the check matrix of $\cal C$ can be easily established
if we write out the check relations $A'p'(c)^T=0$ and $A''p''(c)^T=0$.

Using the construction (\ref{indC}) and
taking ${\cal C}^{0,1}$ and ${\cal C}^{1,0}$ as a base,
we can inductively construct
the class of all codes $\{{\cal C}^{r_1,r_2}\}$.

\def\refname{References%
\renewcommand{\thefootnote}{\dag}%
\footnote{ The references marked by \dag\ have been included during the translation, see Section~\ref{s:0}.}
}
\providecommand\href[2]{#2} \providecommand\url[1]{\href{#1}{#1}}

\end{document}